\documentclass[12pt]{article}

\usepackage{a4}
\usepackage[latin1]{inputenc}
\usepackage[francais]{babel}
\usepackage[all]{xy}
\usepackage{marvosym}
\usepackage{graphicx}
\usepackage{fancybox}
\usepackage{natbib}

\newcommand{\Al}{\mbox{\footnotesize \textsf{A}}}
\newcommand{\Be}{\mbox{\footnotesize \textsf{B}}}

\newcommand{\euro}{\textrm{~\EUR}}

\title{Les crashs sont rationnels}
\author{Pierre Lescanne\\
Université de Lyon, École normale supérieure de Lyon, CNRS (LIP), \\ 46 all\'ee
d'Italie, 69364 Lyon, France}

\begin{document}
\maketitle

  \hrule

\begin{abstract}

  Comme nous le montrons en utilisant les notions d'équilibre dans la théorie des jeux
  séquentiels infinis, les crashs ou les escalades financières sont rationnels, pour les
  agents économiques ou environnementaux, qui ont une vision d'un monde infini.  Cela va à
  l'encontre de la vision auto-régulatrice d'une économie sage et pacifique en équilibre.
  Autrement dit, dans ce contexte, équilibre n'est plus synonyme de stabilité.  Nous tentons de tirer de ce constat quelques
  conséquences méthodologiques et quelques nouveaux modes de pensée que l'on doit adopter,
  notamment en micro\-éco\-no\-mie.  Au kit  de raisonnement (le \og mindware\fg{} des psychologues
  cogniticiens) on doit ajouter la coinduction qui est à la
  base de notre raisonnement sur les jeux infinis et la notion de rationalité doit être raffinée.

\medskip

\centerline{\textbf{Abstract}}

As we show by using notions of equilibrium in infinite sequential games, crashes or
financial escalations are rational for economic or environmental agents, who have a vision
of an infinite world.  This contradicts a picture of a self-regulating, wise and pacific
economy. In other words, in this context, equilibrium is not synonymous of stability.  We try to draw, from this
statement, methodological consequences and new ways of thinking, especially in
microeconomics. To the mindware of the specialists of cognitive psychology, one must add,
coinduction which  is the basis of our reasoning in
infinite games and the notion of rationality must be refined.

\medskip

\noindent \textbf{Mots-clés:} jeu économique, jeu infini, jeu séquentiel, crash, escalade,
microéconomie, bulle spéculative, induction, coinduction.

\medskip

\end{abstract}


\hrule

\newpage{4}

\hfill\parbox{9cm}{
Il y a toujours un peu de folie dans l'amour. 

Mais il y a toujours un peu de raison dans la folie.

\rightline{\emph{Friedrich Nietzsche}}
\rightline{\textsf{Ainsi parlait Zarathoustra} (1898)}
}

\bigskip

Traders spéculant sans borne, pays endettés, crashs en séries, banqueroutes, 
consommation échevelée d'énergie.  Nos modèles d'une économie auto-régulatrice où les
crashs sont improbables sont-ils adaptés? 
Sommes-nous sûrs de bien comprendre le fonctionnement des agents dans un système?   

La crise des subprimes de 2008 a mis en valeur des facteurs qui ont au moins deux
origines. D'une part, elle n'est pas liée aux acteurs, mais à l'interaction entre ces
mêmes acteurs. D'autre
part les outils actuels de la science économique, fondés sur des concepts élaborés vers le
milieu du siècle précédent, ne tiennent pas compte des faits et tendent à plaquer des axiomes et des
méthodes sans confrontation à la réalité.  Cette crise est donc systémique, en ce sens
qu'elle met en cause le système et non les acteurs qui se sont comportés de manière
rationnelle, ayant foi en une croissance infinie. Elle est aussi méthodologique et révèle
une difficulté des économistes à prendre en compte certains mécanismes.





Nous allons ici nous focaliser sur un phénomène bien connu, celui de
l'escalade\,\footnote{C'est un thème de la littérature, de Macbeth à Madame Bovary.}, car,
réfuté comme paradoxal, par les tenants de la rationalité des agents, il démontre
clairement l'inadaptation des principes de raisonnement classiques.  L'escalade consiste à
prendre, sans se donner de borne, une série de décisions de plus en plus lourdes de
conséquences.  Cette fuite en avant qui frappe aujourd'hui aussi bien l'économie et la finance
que le développement social et environnemental est une caractéristique des bulles spéculatives.  Cette apparente irrationalité a été illustrée
par Newton après l'avénement  de l'une des premières crise financières, celle de la
Compagnie des Mers du Sud quand il a déclaré:
\begin{quote}
  \og Je sais calculer le mouvement des planètes, mais je ne sais pas calculer la folie
  des hommes.\fg{}
\end{quote}

Car dans l'escalade, les agents ont un comportement tout ce qu'il y a de plus rationnel, à
condition qu'ils croient en l'inépuisabilité des ressources, qu'elles soient énergétiques,
naturelles ou financières.  En effet, le trader ou le financier est rationnel parce qu'il
raisonne dans son monde qu'il croit infini et parce qu'il pense pouvoir créer de la
monnaie sans fin.  Ce qui est curieux, c'est que l'escaladeur peut, de bonne foi, faire
monter indéfiniment l'enchère ou l'enjeu quitte à perdre de plus en plus.  Pour un
observateur externe, le décideur impliqué dans une escalade semble agir contre le bon
sens, mais du point de vue du décideur et dans son monde fermé, il est tout à fait
raisonnable, comme cela sera démontré par un raisonnement qui implique la mise en
{\oe}uvre de méthodes subtiles de raisonnement sur l'infini.  Cette différence de niveau
de perception, est probablement ce qui distingue la rationalité instrumentale de la
rationalité épistémique (voir section~\ref{sec:esc-psy}).    Par conséquent, celui qui capte à la fois la
vision externe et la vision interne et prend en compte l'une et l'autre est épistémiquement rationnel
tandis que l'agent prisonnier de l'intérieur du système n'est qu'instrumentalement
rationnel.   Le premier verra le crash arriver, alors que le second sera aveugle. 

Enfin, nous ne disons pas que l'escalade et le crash qui peut en être la conséquence sont
inéluctables, mais nous affirmons qu'ils sont plausibles parce que l'escalade est
sous-tendue par une démarche rationnelle des agents.

\section{Vers de nouveaux outils d'analyse  des systèmes}
\hfill\parbox{12cm}{\begin{quote}
  
Il est dans la nature humaine de penser sagement et d'agir d'une façon absurde.

\rightline{\emph{Anatole France}} \rightline{\textsf{Le livre de mon ami}}
\end{quote}}

\bigskip

Dans cet article, nous étudions les systèmes à agents, un système étant une organisation
de taille plus ou moins grande où les entités élémentaires sont appelées des \og agents
\fg{}.  On peut parler d'agent économique ou d'agent écologique.  Dans la suite, nous
emploierons aussi le terme de \og joueur\fg{}, par analogie avec les jeux qui seront notre
paradigme.  En fonction de son intérêt, un agent prend des décisions qui sont les
conséquences de ses préférences ou de l'estimation de ce que ça lui rapporte.  Bien sûr
ces choix ont une influence sur le comportement global du système.  L'un de ces
comportements importants qui nous intéresse se manifeste par le fait que le système est en
\emph{équilibre}, les décisions des agents étant faites pour maximiser les préférences de
chaque agent.  En ce sens, équilibre est synonyme de \emph{stabilité}, mais, comme nous le
verrons, cela peut aussi coïncider avec une évolution rapide des principaux paramètres du
système\,\footnote{En relativité générale, un trou noir est le résultat d'un équilibre
  alors que l'on sait combien extrême est son comportement.}, comme par exemple la hausse
ou la baisse rapide et forte d'un prix.  Ainsi, l'équilibre peut résulter en une grande
instabilité des paramètres majeurs du système, comme c'est le cas dans l'escalade, qui est
le concept sur lequel nous allons nous focaliser.  En gros nous envisageons une suite
\emph{équilibre-décision, équilibre-décision, ...}.  Dans les jeux infinis, équilibre
n'est donc plus synonyme de stabilité.  Cette propension des agents à l'optimisation est
ce que l'on appelle leur \og rationalité\fg{}, autrement dit les agents sont doués de
raison et l'utilisent à leur avantage dans leurs décisions.  Mais la rationalité a deux
facettes, suivant qu'on la mesure de l'extérieur ou de l'intérieur, c'est-à-dire suivant
qu'on en a une perception globale ou locale.  Ces deux points de vue peuvent donner lieu à
des constatations complètement opposées.  L'escalade, phénomène particulièrement
irrationnel d'un point de vue holistique, ne l'est plus d'un point de vue réductionniste.
Dans ce phénomène, plusieurs paradoxes apparaissaient.  Premièrement, une rationalité
locale, celle des agents pris individuellement, peut résulter en une totale irrationalité
quand on l'observe au niveau du système pris dans son entier.  Comme l'agent est coincé
dans son monde, il sera difficile pour un observateur externe de convaincre celui-ci qu'il
se trompe.  Et quand l'observateur affirmera la rationalité ou l'irrationalité des agents,
il devra prendre en compte le niveau où il se place.  Deuxièmement un système fondé sur
des équilibres peut être chaotique et ainsi le crash peut être la conséquence d'un
équilibre.  Donc \og agents rationnels en équilibre au moment de leurs décisions\fg{}, ne
veut pas dire, système à évolution lente et régulière.

Les phénomènes de chaos ou d'escalade sont mal compris dans les disciplines scientifiques
(nous pensons surtout à la microéconomie), qui basent une partie de leur explication sur
les systèmes multi-agents.  Comme Jean-Philippe Bouchaud qui affirme que \og les sciences
économiques ont besoin d'une révolution scientifique\fg{}\,\footnote{J-Ph. Bouchaud.
  \textsf{Economics needs a scientific revolution}.  \emph{Nature}, 455:\penalty0 1181,
  oct 2008.} ou David Collander\,\footnote{cité par Christian Chavagneux \emph{Une brève
    histoire des crises financières: des tulipes aux subprimes}, \textsf{La Découverte},
  Paris, p.~9.} qu'il nous faudra cent cinquante ans pour parvenir à prouver que les faits
que nous vivons depuis 2007 sont possibles en théorie, nous disons qu'elles requièrent une
profonde remise en cause de leurs fondements, de leurs méthodes et de leurs outils.  Parmi
ces nouveaux outils il y a peut-être ceux de la logique comme elle est développée par une
autre grande science des systèmes qu'est l'informatique, au premier rang duquel se trouve
bien sûr la coinduction dont nous allons parler.


\section{Un cas d'école: McDonald's Restaurants contre Morris \& Steel}
\label{sec:un-cas-decole}

\hfill\parbox{9cm}{
 Dis-moi ce que tu manges, je te dirai ce que tu es.

\rightline{\emph{Anthelme Brillat-Savarin,}} 

\rightline{\textsf{La physiologie du goût}}
}

\bigskip

Un cas intéressant d'escalade est celui qui opposa la chaîne de restaurants McDonald's à
deux militants écologistes anglais, David Morris et Helen Steel\,\footnote{John Vidal,
  \textsf{McLibel, Burger Culture on Trial}, \emph{Macmillan Publishers}, 1997}.
Clairement le management de McDonald's a une politique réfléchie, sait argumenter et
s'entoure des meilleurs juristes.  Le procès, qui dura dix ans, fut le plus long procès de
l'histoire juridique britannique.  Les faits commencèrent par une distribution de tracts
accusant la chaîne McDonald's de plusieurs méfaits, pas tous pleinement justifiés et
étayés. S'entourant des stars du barreau londonien et s'appuyant sur le critère de la
menace crédible (dont nous parlerons plus loin), c'est-à-dire le fait que le puissant
McDonald's fasse peur et sur l'idée qu'une compagnie solide comme McDonald's puisse
supporter un procès sans fin\,\footnote{L'auteur a pu mettre en évidence dans le cadre d'un
  contentieux de consommation la même attitude de la part d'un grand leader français de la
  distribution de produits ménagers et électroniques. Malheureusement pour cette
  entreprise une loi récente avait assoupli la procédure de la part du consommateur et
  l'entreprise l'ignorait ou tout au moins pensait que le consommateur ne la connaissait
  pas.}, la compagnie attaqua les militants en diffamation.  Mais deux de ceux-ci
décidèrent de se défendre, sur les bases du \og rien à perdre\fg{}, d'une \og justice
artisanale\fg{} et du \og vous ne me faites pas peur\fg{}.  À la fin d'une longue, très longue,
procédure, ils furent condamnés, car certaines des déclarations des tracts n'ont pas pu
être prouvées comme non diffamatoires. Mais c'est de fait la compagnie qui a perdu.  En
effet, dans le long mémoire du juge, l'image de marque de McDonald's a été plus
qu'écornée. Il a été démontré que les reproches faits par les écologistes sur les
conditions de travail, sur l'exploitation des enfants et sur la cruauté envers les animaux
étaient fondés. Ceci a été mis en évidence à l'audience dans un grand tapage médiatique,
relayé mondialement.
De plus, les militants ont réussi à faire condamner l'État britannique par la Cour
européenne des Droits de l'Homme, sur l'argument que la loi britannique ne permettait pas
au citoyen de faire des griefs à des grandes compagnies multinationales. Ceci aboutit à un
changement de la loi en faveur du citoyen et à des dommages aux requérants. On estime que
pour un dédommagement de 60~000~£ (montant que la Cour européenne des Droits de l'Homme
trouva d'ailleurs exagéré et que McDonald's n'a pas réclamé), la compagnie McDonald's a
perdu 10~000~000~£.  Nous sommes typiquement dans un cas d'escalade.  Bien que la
compagnie vit qu'elle s'enferrait et perdait en image et en frais de procédure, elle ne
put accepter la défaite devant la cour de justice et continua donc à argumenter.  Il est
intéressant de noter qu'un débat préliminaire à l'audience a porté sur la rationalité du
processus de décision, la question étant de savoir si on se référerait à la décision d'un
juge ou à celle d'un jury. Finalement le jury a été rejeté car plus émotionnel dans ses
comportements au profit d'un juge plus rationnel.

\section{Les équilibres dans les jeux}

Dans tous les processus d'escalade il y a un mécanisme d'interaction -- compéti\-tion et
c'est la compétition qui prime.  Très tôt, les philosophes se sont rendus compte que les
règles qui gouvernent les activités des groupes d'individus suivent celles des jeux.
S'est alors développée la \emph{théorie des jeux}.  En effet, comme dans un jeu de
société, les hommes coopèrent plus ou moins, mais surtout agissent pour leur propre
compte.  Très vite est donc apparu le concept de jeu pour décrire comment fonctionnait
l'interaction entre les acteurs d'un système, notamment dans les phénomènes économiques.
L'acte fondateur de la théorie des jeux est le livre de John von Neumann et Oskar
Morgenstern, \emph{Théorie des jeux et comportements économiques}, publié en 1944. 

Jean-Jacques Rousseau explique dans son \emph{Discours sur l'origine et les fondements de
  l'inégalité parmi les hommes} (1775) comment les hommes même acculturés participent à une
activité commune en mêlant à la fois une forme d'interaction et un certain égoïsme qui
les pousse à choisir une option qui correspond le plus à leur intérêt immédiat.  L'exemple
qu'il choisit est celui de la chasse au cerf et cet exemple est resté célèbre.

\begin{quote}
  \og Voilà comment les hommes purent insensiblement acquérir quelque idée grossière des
  engagements mutuels, et de l'avantage de les remplir, mais seulement autant que pouvait
  l'exiger l'intérêt présent et sensible ; car la prévoyance n'était rien pour eux, et
  loin de s'occuper d'un avenir éloigné, ils ne songeaient pas même au
  lendemain. S'agissait-il de prendre un cerf, chacun sentait bien qu'il devait pour cela
  garder fidèlement son poste ; mais si un lièvre venait à passer à la portée de l'un
  d'eux, il ne faut pas douter qu'il ne le poursuivît sans scrupule, et qu'ayant atteint
  sa proie il ne se souciât fort peu de faire manquer la leur à ses compagnons.\fg{}
\end{quote}

Dans cette dualité, intérêt collectif -- intérêt individuel, l'homme adopte une ligne de
conduite de laquelle il n'a aucun intérêt à dévier.  Une position stratégique, où les
comportements individuels sont bloqués par le fait que les acteurs ne changent plus leurs
choix, est appelée un \emph{équilibre}.  Dans son ouvrage \emph{Recherches sur les
  principes mathématiques de la théorie des richesses} (1838), l'économiste
Antoine-Augustin Cournot met en évidence, dans le cas des duopoles, cette notion
d'équilibre, qui n'a pas encore son nom.  Deux entreprises sont en compétition pour
produire les mêmes objets, elles essaient d'ajuster leur production pour optimiser leurs
profits.  En calculant les équilibres de Cournot, on peut calculer la quantité optimum que
chaque entreprise doit fabriquer pour gagner le plus d'argent.  Une des
caractéristiques des duopoles de Cournot et de la chasse de Rousseau est l'interaction
sans coopération.  Nous allons donc par la suite nous intéresser aux jeux non coopératifs
où chaque joueur est égoïste et ne cherche pas à aider les autres, même si cette aide peut
lui profiter à terme.  En restant dans le cadre des jeux non coopératifs, l'on passe de
Cournot à John F.~Nash qui, en 1947, établit une forme générale d'équilibre que l'on
appelle \emph{équilibre de Nash}, qui a eu pour effet d'initier une recherche active sur les
jeux non coopératifs.  Ces équilibres sont définis sur une forme particulière de jeux que
l'on appelle les \emph{jeux en forme normale}. Ce sont les jeux qui se font en un seul
coup et où les gains sont immédiatement rapportés.  Un exemple  typique de jeu en forme normale est le
jeu \emph{pierre-papier-ciseaux}.  Deux joueurs dévoilent au même moment l'un des trois
objets suivants, soit une pierre, soit une feuille de papier, soit une paire de ciseaux.
Les règles sont les suivantes:
\begin{itemize}
\item le papier l'emporte sur la pierre,
\item la pierre l'emporte sur les ciseaux,
\item les ciseaux l'emportent sur le papier.
\end{itemize}
Dans ce cas, chaque joueur doit probabiliser ses choix et sa stratégie optimum est de
jouer $1/3$ des fois papier, $1/3$ des fois pierre et $1/3$ des fois ciseaux.  Le jeu de
l'\emph{appariement des pièces} est une version simplifiée du jeu
\emph{pierre-papier-ciseaux}.  Deux joueurs, que nous appellerons \textsf{Alice} et
\textsf{Bertrand} présentent au même moment une pièce de monnaie, soit côté \emph{pile},
soit côté
\emph{face}.  La convention est suivante:
\begin{itemize}
\item si les deux joueurs choisissent pile ou si les deux joueurs choisissent face, alors c'est
  \textsf{Alice} qui gagne,
\item si les deux joueurs choisissent des côtés différents, alors c'est \textsf{Bertrand}
  qui gagne.
\end{itemize}
Là encore les deux joueurs ont intérêt à présenter pile ou face avec une égale
probabilité.  Nous ne nous étendrons pas trop sur ce type de jeux, car il y a de notre
point de vue trop d'arbitraire attribué à la valeur pécuniaire des gains, pourtant au c{\oe}ur du calcul des
probabilités et nous affirmons que les acteurs jouent sans quantifier leurs gains.  Les jeux
qui nous intéressent sont les jeux séquentiels non probabilistes, proposés par Harold Kuhn
en 1953, et dans la suite les sommes allouées seront symboliques et n'auront pas valeur de
nombre. Seul compte pour nous le fait que l'on puisse comparer des entités.

\section{Les jeux séquentiels}
\label{sec:les-jeux-sequentiels}
\hfill\parbox{6cm}{\begin{quote}
Je te tiens, tu me tiens 

par la barbichette.

\rightline{\emph{Comptine}}
\end{quote}}

\bigskip

Dans un jeu séquentiel, les joueurs jouent plusieurs fois, chacun à leur tour, la
répartition des gains se faisant à la fin du jeu.  Dans ces jeux, il y a plusieurs sortes
d'équilibres, mais ceux qui nous intéressent vraiment sont les équilibres dits
d'\emph{induction rétrograde}, parce qu'ils traduisent la rationalité des choix des
joueurs.  Ils sont appelés \og rétrogrades\fg{}, parce qu'ils se calculent en marche
arrière, en partant de la fin du jeu.  Sans nuire à la généralité, nous considérerons
seulement les jeux à deux joueurs.  Par exemple, considérons une variante du jeu de
l'\emph{appariement des pièces}, où cette fois-ci les joueurs, que nous nommerons
\textsf{Alice} et \textsf{Bertrand}, jouent en séquence.  Plus précisément, \textsf{Alice}
joue d'abord, puis \textsf{Bertrand}, puis \textsf{Alice} à nouveau.  En cas,
d'appariement (deux piles consécutifs ou deux faces consécutives) \textsf{Alice} gagne un
point.  Dans le cas contraire, \textsf{Bertrand} gagne un point.  Les points sont
accumulés et distribués à la fin de la partie.  Par exemple si \textsf{Alice} joue pile,
puis \textsf{Bertrand} joue face, puis \textsf{Alice} joue pile, il n'y a pas
d'appariement, donc \textsf{Alice} ne gagne rien et \textsf{Bertrand} gagne les deux
points.  Le gain final est $0$ pour \textsf{Alice} et $2$ pour \textsf{Bertrand}.

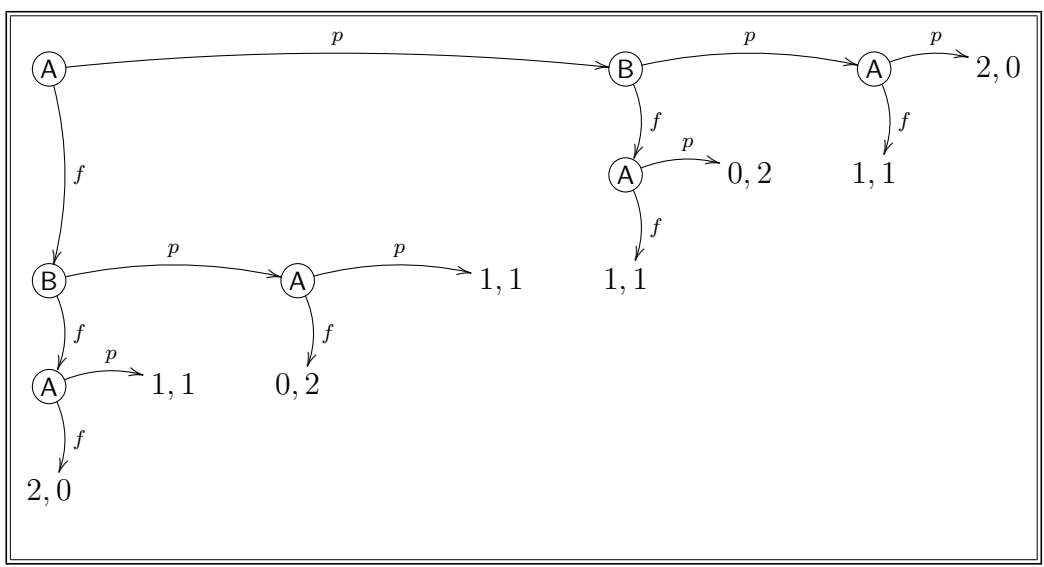
\begin{figure}[htb!]
  \centering \doublebox{\parbox{.85\textwidth}{
\begin{center}
    \xymatrix{ 
      *+[o][F]{\Al} \ar@/^/[rrrrr]^p \ar@/^/[dd]^f &&&&& *+[o][F]{\Be}
      \ar@/^/[rr]^p \ar@/^/[d]^f &&*+[o][F]{\Al} \ar@/^/[d]^f\ar@/^/[r]^p &
      2,0\\
      &&&&&*+[o][F]{\Al}\ar@/^/[d]^f\ar@/^/[r]^p&0,2&1,1\\
      *+[o][F]{\Be}\ar@/^/[rr]^p \ar@/^/[d]^f && *+[o][F]{\Al} \ar@/^/[rr]^p\ar@/^/[d]^f&& 1,1 &1,1&&\\
      *+[o][F]{\Al}\ar@/^/[d]^f \ar@/^/[r]^p & 1,1 & 0,2\\
      2,0 }
  \end{center}
}}
\caption{Le jeu séquentiel de l'appariement de la pièce, où $\Al$ commence, puis $\Be$~joue,
  puis $\Al$ joue à nouveau.}
\label{fig:appar}
\end{figure}

Ce jeu possède une stratégie gagnante presqu'évidente, mais il n'est là qu'à titre
d'illustration, pour montrer qu'il faut considérer les stratégies même les plus stupides, en
vertu de la contrafactualité dont nous parlerons plus loin. 
Très naturellement, ce jeu se représente par
le diagramme de la figure~\ref{fig:appar} où \textsf{Alice} est identifiée par \Al{} et
\textsf{Bertrand} est identifié par \Be{}, et où les flèches sont les étapes de jeu.
L'étiquette $p$ d'une flèche indique que le joueur a joué pile tandis que l'étiquette $f$
indique que le joueur a joué face.  Insistons sur le fait que là encore, la valeur donnée
aux gains n'a pas de sens hormis le fait qu'ils peuvent être comparés; ainsi $0 < 1$ et $1
<2$.  Le nombre $0$ veut dire que le joueur n'a jamais gagné, le nombre $1$ veut dire
qu'il a gagné autant de fois que son adversaire et le nombre $2$ veut dire qu'il a gagné
plus de fois que son adversaire.

Sur le diagramme, chaque joueur a deux choix: soit aller en bas (face), soit aller à
droite (pile).  Partant du point le plus haut et le plus à gauche, en suivant la flèche
étiquetée $p$, puis la flèche étiquetée  $f$, puis la flèche étiquetée $p$, on
arrive à la fin du jeu. L'inscription $0,2$ signifie  qu'\og \textsf{Alice} gagne $0$
fois et \textsf{Bertrand} gagne $2$ fois\fg{}.  Appelons $p\,f\,p$ cette ligne de jeu.

\begin{figure}[!hbp]
  \centering
     \doublebox{\parbox{.85\textwidth}{
   \begin{center}
    \xymatrix@R=10pt@C=12pt{
      *+[o][F]{\Al} \ar@{=>}@/^/[rrrrr]^p \ar@/^/[dd]^f &&&&&
      *+[o][F]{\Be} \ar@/^/[rr]^p \ar@{=>}@/^/[d]^f &&
      *+[o][F]{\Al}\ar@/^/[d]^f\ar@{=>}@/^/[r]^p &
      2,0\\
      &&&&&*+[o][F]{\Al}\ar@{=>}@/^/[d]^f\ar@/^/[r]^p&0,2&1,1\\
      *+[o][F]{\Be}\ar@{=>}@/^/[rr]^p \ar@/^/[d]^f 
      && *+[o][F]{\Al} \ar@{=>}@/^/[rr]^p\ar@/^/[d]^f&& 1,1 &1,1&&\\
      *+[o][F]{\Al}\ar@{=>}@/^/[d]^f \ar@/^/[r]^p & 1,1 & 0,2\\
      2,0 }
  \end{center}
  \begin{center}
    \xymatrix@R=10pt@C=12pt{
      *+[o][F]{\Al} \ar@/^/[rrrrr]^p \ar@{=>}@/^/[dd]^f &&&&&
      *+[o][F]{\Be} \ar@/^/[rr]^p \ar@{=>}@/^/[d]^f &&
      *+[o][F]{\Al}\ar@/^/[d]^f\ar@{=>}@/^/[r]^p &
      2,0\\
      &&&&&*+[o][F]{\Al}\ar@{=>}@/^/[d]^f\ar@/^/[r]^p&0,2&1,1\\
      *+[o][F]{\Be}\ar@{=>}@/^/[rr]^p \ar@/^/[d]^f 
      && *+[o][F]{\Al} \ar@{=>}@/^/[rr]^p\ar@/^/[d]^f&& 1,1 &1,1&&\\
      *+[o][F]{\Al}\ar@{=>}@/^/[d]^f \ar@/^/[r]^p & 1,1 & 0,2\\
      2,0 }
  \end{center}
}}
  \caption{Les deux équilibres du jeu de l'appariement des pièces.  \emph{ Les doubles
      flèches correspondent aux choix faits par les joueurs}}
\label{fig:4eq}
\end{figure}
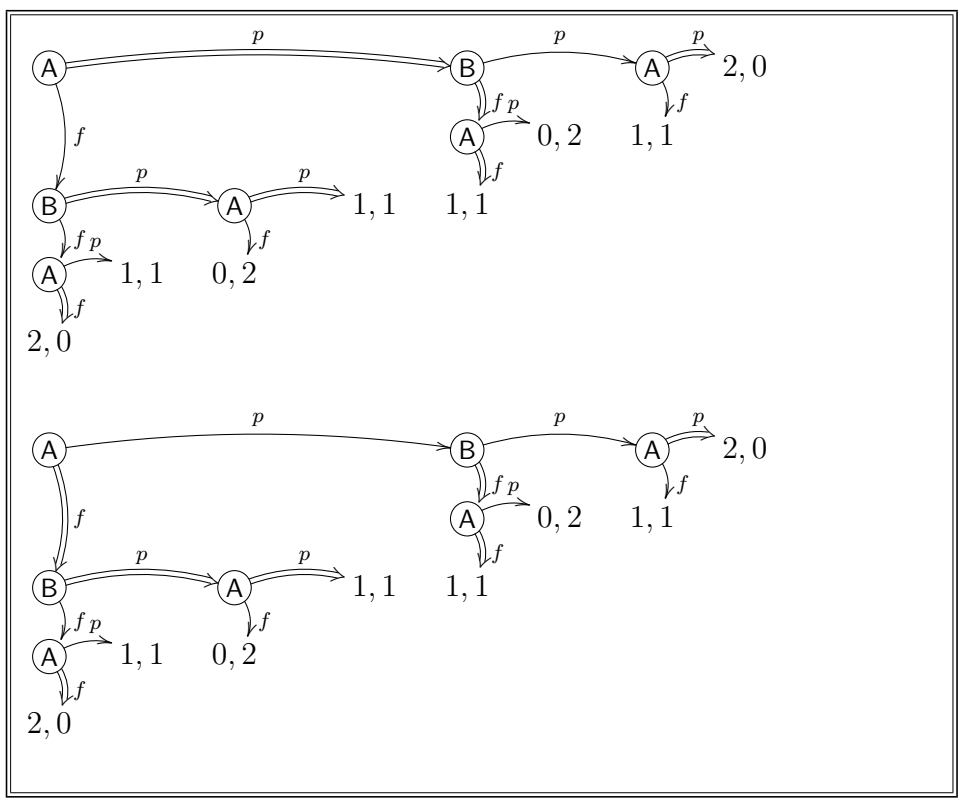

Dans ce jeu, parmi les huit lignes de jeu, deux sont des équilibres (Fig.~\ref{fig:4eq}).
Ansi dans la ligne de jeu, $p\,f\,f$ qui aboutit à $1,1$, si \textsf{Bertrand} change son
choix en~$p$, alors \textsf{Alice} changera son choix en $p$ et il gagnera $0$ fois; donc
il ne le fait pas.  Le calcul d'un tel équilibre fait appel à ce que les spécialistes de
théorie des jeux appellent une \emph{induction rétrograde}, que les logiciens appellent
tout simplement une \emph{induction} ou une \emph{récurrence}.  Prenons une position du
jeu et considérons les positions qui viennent après (en suivant les flèches). Si l'on
regarde de près cette configuration, on constate que c'est aussi un jeu, nous dirons un
sous-jeu, car il est complètement inclus dans le jeu de la figure~\ref{fig:appar}.  L'idée
est d'attribuer à chaque sous-jeu un couple de gains qui correspond à ce que retourne
l'équilibre.  Pour associer ces couples on part des fins de parties et on remonte vers le
début de la partie.  Peu à peu, on construit les affectations de gains pour les
sous-jeux. On part des fin de jeux, puis on construit des jeux un peu plus grands, puis
plus grands encore, pour arriver au jeu le plus grand, c'est à dire le jeu qui nous
intéresse. Autrement dit, on construit les équilibres pas à pas.  A chaque fois, on
choisit pour le joueur dont c'est le tour, le plus grand des gains, et on retient le
choix.  L'équilibre est la ligne de jeu qui passe par les positions retenues.

Une caractéristique de l'analyse des jeux séquentiels est la
contrafactualité\,\footnote{Un raisonnement est contrafactuel s'il s'appuie sur des
  hypothèses qui n'ont pas besoin d'être plausibles. \og S'il y a des martiens,
  ils chercheront à communiquer avec nous.\fg{}}.  On raisonne dans toutes les situations comme si
elles étaient possibles. Or clairement, certaines positions de jeu, situées après que l'un
des joueurs a arrêté, ne se produiront pas, mais elles doivent être prises en compte pour
expliquer comment les joueurs raisonnent.  La contrafactualité joue un rôle encore plus
important dans le cas des jeux infinis si cruciaux dans l'escalade.

\paragraph{Quel lien entre les équilibres rétrogrades et la rationalité?}
\label{sec:quel-lien-entre}

Robert Aumann, prix Nobel d'économie en 2005, a démontré en 1995 que les équilibres
rétrogrades sont les lignes de jeu rationnelles.  Ce sont les lignes de jeu, où tout le
monde sait que personne ne peut changer ses choix sans y perdre et tout le monde sait
qu'il ne peut pas, en toute rationalité, en être autrement.  On dit que les joueurs
choisissent ces équilibres à partir de la connaissance commune qu'ils ont de la
rationalité des autres joueurs.

\section{Les jeux 0,1}
\label{sec:le-jeu-0}
\hfill\parbox{6cm}{\begin{quote}
Vulnerant omnes

Ultima necat

\rightline{\emph{Phrase placée sur les cadrans solaires,}}
\end{quote}}

\bigskip

Nous allons nous intéresser à des jeux séquentiels très simples.  Dans ces jeux, les joueurs ont le
choix d'abandonner ($a$) ou de continuer ($c$). Les gains
sont $0\!\euro$ ou~$1\!\euro$\,\footnote{Pour pimenter le jeu, on pourrait remplacer $1\!\euro$ par
$1~000~000~000\!\euro$ et donc dire que les joueurs, soit ne reçoivent rien, soit reçoivent un
milliard d'euros. Nous insistons sur le fait que la quantité ne compte pas, seule compte la comparaison.}.
Concentrons-nous sur le jeu à sept tours.  

\begin{displaymath}
  \xymatrix{
*++[o][F]{\Al} \ar@/^/[r]^c \ar@/^/[d]^a &*++[o][F]{\Be} \ar@/^/[r]^c \ar@/^/[d]^a &*++[o][F]{\Al} \ar@/^/[r]^c \ar@/^/[d]^a &*++[o][F]{\Be} \ar@/^/[r]^c \ar@/^/[d]^a &*++[o][F]{\Al} \ar@/^/[r]^c \ar@/^/[d]^a &*++[o][F]{\Be} \ar@/^/[r]^c \ar@/^/[d]^a &*++[o][F]{\Al} \ar@/^/[r]^c \ar@/^/[d]^a &1,0\\
0,1&1,0&0,1&1,0&0,1&1,0&0,1
}
\end{displaymath}

\begin{itemize}
\item Au dernier tour, \textsf{Alice} doit bien sûr choisir $c$, car elle gagne $1\!\euro$.
\item A l'avant dernier tour, \textsf{Bertrand} a le choix, pauvre choix ! En effet, s'il continue il ne
  gagne rien et s'il abandonne il ne gagne rien non plus.
\item Au cinquième tour, \textsf{Alice} continue car dans tous les cas, elle gagne $1\!\euro$
  et si elle abandonne, elle ne gagne  rien du tout.
\item Au quatrième tour, \textsf{Bertrand} a le même choix qu'à l'avant dernier tour,
  à savoir ne rien gagner ou ne rien gagner.
\item Au troisième tour, \textsf{Alice} continue.
\item Au deuxième tour, \textsf{Bertrand} a le choix entre rien et rien.
\item Au début du jeu, \textsf{Alice} ne s'arrête surtout pas. 
\end{itemize}

Le jeu $0,1$ à sept tours a donc plusieurs lignes de jeu rationnelles, qui ont la caractéristique
qu'\textsf{Alice} continue toujours et \textsf{Bertrand} fait ce qu'il veut.   Considérons maintenant
le jeu $0,1$ à six tours.

\begin{displaymath}
  \xymatrix{
*++[o][F]{\Al} \ar@/^/[r]^c \ar@/^/[d]^a &*++[o][F]{\Be} \ar@/^/[r]^c \ar@/^/[d]^a &*++[o][F]{\Al} \ar@/^/[r]^c \ar@/^/[d]^a &*++[o][F]{\Be} \ar@/^/[r]^c \ar@/^/[d]^a &*++[o][F]{\Al} \ar@/^/[r]^c \ar@/^/[d]^a &*++[o][F]{\Be} \ar@/^/[r]^c \ar@/^/[d]^a &0,1\\
0,1&1,0&0,1&1,0&0,1&1,0
}
\end{displaymath}

Cette fois-ci les choses changent pour l'équilibre d'induction rétrograde.
\begin{itemize}
\item Au dernier tour, \textsf{Bertrand} continue.
\item A l'avant dernier et cinquième tour, \textsf{Alice} fait ce qu'elle veut.
\item Au quatrième tour \textsf{Bertrand} continue,
\item Au troisième tour, \textsf{Alice} fait ce qu'elle veut.
\item Etc.
\end{itemize}
Voici le diagramme de celui où \textsf{Alice} abandonne toujours:
\begin{displaymath}
  \xymatrix{
*++[o][F]{\Al} \ar@/^/[r]^c \ar@{=>}@/^/[d]^a &*++[o][F]{\Be} \ar@{=>}@/^/[r]^c \ar@/^/[d]^a
&*++[o][F]{\Al} \ar@/^/[r]^c \ar@{=>}@/^/[d]^a 
&*++[o][F]{\Be} \ar@{=>}@/^/[r]^c \ar@/^/[d]^a &*++[o][F]{\Al} \ar@/^/[r]^c \ar@{=>}@/^/[d]^a 
&*++[o][F]{\Be} \ar@{=>}@/^/[r]^c \ar@/^/[d]^a &0,1\\
0,1&1,0&0,1&1,0&0,1&1,0
}
\end{displaymath}

Dans ce jeu, et dans tous les jeux $0,1$ à nombre pair de tours, les lignes de jeu
rationelles, c'est-à-dire, les équilibres se situent quand \textsf{Bertrand} continue et \textsf{Alice}
fait ce qu'elle veut.  Nous examinerons des jeux $0,1$  infinies et nous verrons quels
sont les équilibres dans ces jeux.

\section{L'enchère à l'américaine}
\label{sec:lenchere-americaine}
\hfill\parbox{12cm}{\begin{quote}
\hspace*{8cm}{tous deux}

Marchent droit l'un vers l'autre, et le duel recommence.

Voilà que par degrés de sa sombre démence

Le combat les enivre ; il leur revient au c{\oe}ur

Ce je ne sais quel dieu qui veut qu'on soit vainqueur,

\rightline{\emph{Victor Hugo,}} \rightline{\textsf{La Légende des siècles. Le mariage de Roland}}
\end{quote}}

\bigskip

Ce jeu a été décrit par le chercheur américain Martin Shubik\,\footnote{M.~Shubik, \textsf{The
  dollar auction game: A paradox in noncooperative behavior and escalation}. \emph{Journal
    of Conflict Resolution}, 15\penalty0 (1):\penalty0 109--111, 1971.} en 1971, mais il
est bien connu en France, sous le nom d'\emph{enchère à l'américaine}.  Dans notre pays,
il se pratique, par exemple, dans certains mariages et consiste à mettre aux enchères la
jarretière de la mariée, le produit de la cagnotte récoltée servant aux jeunes époux à se
payer leur voyage de noce ou à s'installer.  Dans sa version, Shubik met aux enchères un
dollar, mais le but est bien sûr de récolter beaucoup plus que le prix de l'objet.
L'enchère à l'américaine consiste à \og mettre en vente\fg{} l'objet de la manière
suivante: chaque fois qu'une personne surenchérit d'un montant de $n\!\euro$, elle doit
mettre la somme en question dans un chapeau et cette somme ne lui revient pas.  Ainsi que
l'écrit Shubik, il est conseillé d'attendre pour démarrer l'enchère que l'ambiance soit
bien échauffée, comme c'est le cas à la fin d'une noce et l'on peut se limiter à deux
enchérisseurs, car cela ne change pas le phénomène, qui de toute façon la plupart du
temps, fait se terminer l'enchère avec deux participants qui ne veulent pas lâcher le
morceau.  C'est alors que l'on constate un phénomène d'escalade, les enchérisseurs vont
payer plus que la valeur de l'objet et plus que ce qu'ils pensaient investir à l'origine
pour continuer à se battre en vue d'acquérir la jarretière ou le billet vert à l'éffigie
de Georges Washington.  En fait, ils ont déjà tant investi qu'ils ne veulent pas
abandonner l'enchère sans récupérer l'object convoité.

Dans l'analyse de Shubik et dans celle de ses successeurs, on note deux affirmations contradictoires.
\begin{itemize}
\item \emph{Pour qu'il y ait escalade, il faut que le jeu soit infini. } Plus précisément,
  Shubik écrit que \og l'analyse doit être restreinte aux jeux infinis, qui n'ont pas de fin
  spécifique, car aucun phénomène intéressant n'apparait si une borne supérieure [des
  enchères] est établie\fg{}.
\item \emph{Shubik et ses successeurs analysent des jeux finis}, puis extrapolent
  leur résultat au jeu infini. 
\end{itemize}
De là ils concluent faussement que le seul équilibre du jeu est celui où aucun joueur ne
commence les enchères et n'enchérit jamais.  Le joueur rationnel refuserait donc à tout prix
tout type d'escalade.
Nous avons pu démontrer
que cette ligne de jeu n'est pas un équilibre pour l'enchère à l'américaine.
Extrapoler les jeux finis aux jeux infinis, comme le fait Shubik et ceux qui ont repris
son raisonnement ou des variantes est une erreur. Nous reviendrons sur la
faille de ce mode de raisonnement.

A ce point de la présentation, il faut signaler que quand on raisonne, il y a une
différence entre le très très grand et l'infini.  La question n'est pas de savoir si l'on
se restreint à des nombres qu'on peut imaginer ou concevoir ou même écrire, mais
d'affirmer que les nombres n'ont pas de limite.

\section{Les jeux séquentiels infinis: retour sur le jeu~0,1}
\label{sec:retour-sur-0,1}

\hfill\parbox{8cm}{\begin{quote}
  Il marche dans la plaine immense,

Va, vient, lance la graine au loin,

Rouvre sa main, et recommence,

Et je médite, obscur témoin,

\rightline{\emph{Victor Hugo,}} \rightline{\textsf{Saison des semailles. Le soir}}
\end{quote}}

\bigskip

Dans les jeux $0,1$ nous avons limité le nombre de tours.  Mais rien n'empêche de
considérer une infinité de tours.

\begin{displaymath}
  \xymatrix{
*++[o][F]{\Al} \ar@/^/[r]^c \ar@/^/[d]^a &*++[o][F]{\Be} \ar@/^/[r]^c \ar@/^/[d]^a
&*++[o][F]{\Al} \ar@/^/[r]^c \ar@/^/[d]^a &*++[o][F]{\Be} \ar@/^/[r]^c \ar@/^/[d]^a 
&*++[o][F]{\Al} \ar@/^/[r]^c \ar@/^/[d]^a &*++[o][F]{\Be} \ar@{.>}@/^/[r]^c \ar@/^/[d]^a 
&\ar@{.>}@/^/[r]^c \ar@{.>}@/^/[d]^a &\ar@{.>}@/^/[r]^c&\\
0,1&1,0&0,1&1,0&0,1&1,0&&
}
\end{displaymath}

Quels sont les équilibres dans ce cas?  On ne peut plus définir les équilibres par
induction rétrograde, mais on peut le faire par une méthode très similaire\,\footnote{Les
  spécialistes à la suite de Selten (1965) parlent d'\emph{équilibres parfaits en
    sous-jeu}, mais nous proposons de les appeler \emph{équilibres de coinduction
    rétrograde}.}.  Le raisonnement suppose que l'on raisonne sur un objet infini.  La
base de celui-ci, qui a été conçu pour s'adapter aux objets infinis s'appelle la
\emph{coinduction}.  On fait donc de la \og coinduction rétrograde\fg{}.  En fait, il y a
au moins deux équilibres:
\begin{enumerate}
\item \textsf{Alice} abandonne toujours et \textsf{Bertrand} continue toujours (Fig.~\ref{fig:AaBc}),
   \begin{figure}[hbt!]
  \centering
  \doublebox{\parbox{.85\textwidth}{ \centering \xymatrix{ *++[o][F]{\Al}
        \ar@{->}@/^/[r]^c \ar@{=>}@/^/[d]^a &*++[o][F]{\Be} \ar@{=>}@/^/[r]^c
        \ar@{->}@/^/[d]^a &*++[o][F]{\Al} \ar@{->}@/^/[r]^c \ar@{=>}@/^/[d]^a
        &*++[o][F]{\Be} \ar@{=>}@/^/[r]^c \ar@{->}@/^/[d]^a &*++[o][F]{\Al}
        \ar@{->}@/^/[r]^c \ar@{=>}@/^/[d]^a &*++[o][F]{\Be} \ar@2{.>}@/^/[r]^c
        \ar@{->}@/^/[d]^a &\ar@{.>}@/^/[r]^c \ar@2{.>}@/^/[d]^a
        &\ar@2{.>}@/^/[r]^c&\\
        0,1&1,0&0,1&1,0&0,1&1,0&& }
  \caption{\textsf{Alice} abandonne toujours et \textsf{Bertrand} continue
    toujours.}\label{fig:AaBc}
}}
\end{figure} 
\item \textsf{Alice} continue toujours et \textsf{Bertrand} abandonne toujours  (Fig.~\ref{fig:AcBa}).
\begin{figure*}[hbt!]
    \centering
     \doublebox{\parbox{.85\textwidth}{
     \begin{center}
     \xymatrix{ *++[o][F]{\Al} \ar@{=>}@/^/[r]^c \ar@/^/[d]^a &*++[o][F]{\Be}
        \ar@/^/[r]^c \ar@{=>}@/^/[d]^a &*++[o][F]{\Al} \ar@{=>}@/^/[r]^c \ar@/^/[d]^a
        &*++[o][F]{\Be} \ar@/^/[r]^c \ar@{=>}@/^/[d]^a &*++[o][F]{\Al} \ar@{=>}@/^/[r]^c
        \ar@/^/[d]^a &*++[o][F]{\Be} \ar@{.>}@/^/[r]^c \ar@{=>}@/^/[d]^a
        &\ar@2{.>}@/^/[r]^c \ar@{.>}@/^/[d]^a &\ar@{.>}@/^/[r]^c&\\
        0,1&1,0&0,1&1,0&0,1&1,0&& }
    \end{center}
    \caption{\textsf{Alice} continue toujours et \textsf{Bertrand} continue
      toujours.}\label{fig:AcBa}
  }}
   \end{figure*}
\end{enumerate}

\begin{figure}[htb!]
  \centering
  \doublebox{\parbox{\textwidth}{
    \begin{displaymath}
      \xymatrix@C=10pt{
        &\ar@{.>}[r]& *++[o][F]{\Al} \ar@/^1pc/[rr]^c \ar@/^/[d]^a 
        &&*++[o][F]{\Be} \ar@/^1pc/[ll]^c \ar@/^/[d]^a \\
        &&0,1&&1,0
      }
      \qquad 
      \xymatrix@C=10pt{
        &\ar@{.>}[r]& *++[o][F]{\Al} \ar@{=>}@/^1pc/[rr]^c \ar@/^/[d]^a 
        &&*++[o][F]{\Be} \ar@/^1pc/[ll]^c \ar@{=>}@/^/[d]^a \\
        &&0,1&&1,0
      } 
      \qquad
      \xymatrix@C=10pt{
        &\ar@{.>}[r]& *++[o][F]{\Al} \ar@/^1pc/[rr]^c \ar@{=>}@/^/[d]^a 
        &&*++[o][F]{\Be} \ar@{=>}@/^1pc/[ll]^c \ar@/^/[d]^a \\
        &&0,1&&1,0
      }
\end{displaymath}
}}
\caption{Le jeu $0,1$ et ses équilibres vus de façon compacte}
\label{fig:boucle}
\end{figure}
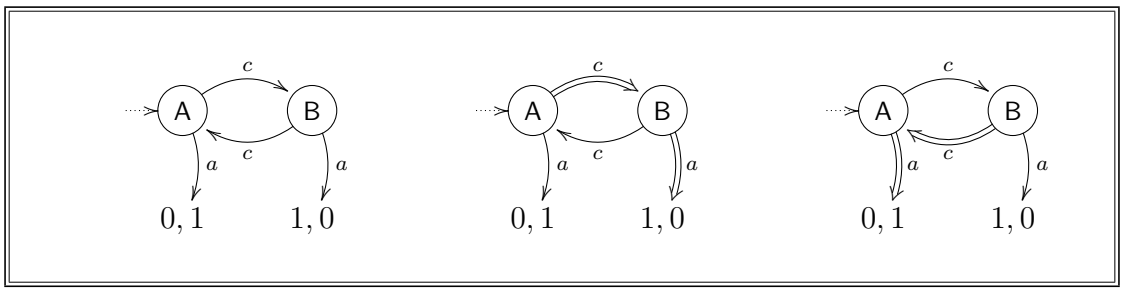

\begin{small}
\paragraph{Justification} \emph{(Ce paragraphe peut être sauté en première lecture)}
\label{sec:justification}
Voyons pourquoi et comment ça marche.  Rappelons comment nous procédons dans le cas des
jeux finis.  Partant des fins de partie, nous examinons une position du jeu.  Il y a un
joueur et deux sous-jeux dont nous connaissons les équilibres; nous choisissons comme
équilibre celui qui correspond au joueur faisant le choix du meilleur des deux équilibres.
Maintenant passons à l'infini et montrons que la ligne où \textsf{Alice} abandonne
toujours et \textsf{Bertrand} continue toujours est un équilibre.  Comme dans le cas de
l'induction rétrograde, nous connaissons les équilibres pour les sous-jeux.  Dans le cas
où c'est le tour de \textsf{Bertrand}, quels sont ces équilibres?  Il y en a deux, l'un
 correspond à l'abandon de \textsf{Bertrand} et dans l'autre, où \textsf{Bertrand}
continue, on a un sous-équilibre où \textsf{Alice} commence. Quel est ce sous-équilibre où
\textsf{Alice} commence?  C'est le même équilibre que celui que l'on cherche, à savoir la ligne de
jeu où \textsf{Alice} abandonne toujours et \textsf{Bertrand} continue toujours!  Des
lignes de jeu optimales que \textsf{Bertrand} a à sa disposition, quel est la meilleure?
Celle où \textsf{Bertrand} continue, puisqu'\textsf{Alice} abandonne et il gagne
$1\!\euro$, tandis que s'il abandonne, il ne gagne rien.  Donc le bon choix, qui donne
l'équilibre, est celui où \textsf{Bertrand} continue, puis \textsf{Alice} abandonne
toujours, puis \textsf{Bertrand} continue toujours.  Dans le cas où c'est le tour
d'\textsf{Alice}, un raisonnement similaire nous montre qu'un équilibre est celui où
\textsf{Alice} abandonne toujours et \textsf{Bertrand} continue toujours.  Résumons nous:
nous avons montré que si on fait l'hypothèse que l'équilibre est la ligne de jeu où
\textsf{Alice} abandonne toujours et \textsf{Bertrand} continue toujours, alors
l'équilibre est la ligne de jeu où \textsf{Alice} abandonne toujours et \textsf{Bertrand}
continue toujours.  Peut-être que cette explication un peu alambiquée se comprend mieux sur la
figure~\ref{fig:boucle} où nous avons représenté le jeu $0,1$ de façon plus compacte.
Nous semblons tourner en rond. Il n'en est rien! Et ce raisonnement est bien correct, car
l'hypothèse porte sur un sous-jeu strict.  Nous l'appellerons \emph{coinduction
  rétrograde}: co-induction vient du nom \emph{induction} et du préfixe \textsf{co}, \og
associé à\fg{} et \textsf{rétrograde} insiste sur l'analogie avec l'induction rétrograde.
En fait, nous avons montré que la ligne de jeu où \textsf{Alice} abandonne toujours et
\textsf{Bertrand} continue toujours est un équilibre tout au long du jeu infini, on dit
que c'est un \og invariant\fg{} du jeu infini.
\end{small}

Le même type de raisonnement s'applique pour l'enchère à l'américaine.  C'est un tout
petit peu plus compliqué, car ce n'est pas le même jeu qui est un invariant, mais le même
jeu paramétré pour tenir compte du fait que les sommes impliquées augmentent.  Il y a donc
au moins deux équilibres dans l'enchère à l'américaine: l'un des deux joueurs continuant
toujours et l'autre abandonnant toujours.

\section{L'escalade est rationnelle}
\label{sec:lesc-est-rati}
\hfill\parbox{12cm}{
L'exubérance irrationnelle a provoqué une escalade excessive des prix des actions.

\rightline{\emph{Alan Greenspan,}} \rightline{\textsf{Président de la Réserve fédérale
    américaine, le 12 mai 1996}}
}

\bigskip

Nous connaissons les équilibres du jeu, $0,1$ infini, mais comment cela conduit-il à une
escalade?  Analysons la situation en faisant l'hypothèse que nos joueurs ou nos acteurs
économiques, sont rationnels mais n'ont pas de mémoire.  Ils n'apprennent rien du passé
qu'ils oublient immédiatement.  Ils savent analyser une situation et comprendre leur
intérêt, mais ne savent pas profiter de leur expérience, ne savent pas se voir raisonner
et ne savent pas se remettre en cause. Ils ne savent pas non plus
évaluer les frais annexes et voir une société qui change. Bref, ils sont complètement
introvertis et oublieux.  A tout moment du jeu, \textsf{Alice} commence un jeu $0,1$ infini. Elle sait
qu'elle a deux stratégies rationnelles possibles: continuer aujourd'hui, demain,
après-demain et toujours en se disant que \textsf{Bertrand} va avoir peur et abandonner
aujourd'hui et toujours dans le futur, ou bien abandonner en pensant que \textsf{Bertrand}
continuera toujours. Dans cette deuxième hypothèse, elle prend très au sérieux la
détermination de \textsf{Bertrand} de continuer toujours, c'est le phénomène de la
\emph{menace crédible} qui est l'attitude des amis de Morris et Steel\,\footnote{Il y
  avait cinq distributeurs de tracts, parmi lesquels seuls Morris et Steel ont affronté
  McDonald's.} face à McDonalds.  \textsf{Alice} continuera si elle pense qu'elle a
impressionné suffisamment \textsf{Bertrand} pour le terroriser et le faire abandonner pour
toujours.  C'est l'attitude de McDonalds tout au long du procès contre Morris et
Steel. \og Ça n'est pas possible, ils vont lâcher prise!\fg{} pensaient ses dirigeants.
\textsf{Bertrand} est dans la même situation et peut décider de continuer.  Si l'on est
dans la situation où aucun des participants ne prend la menace de l'autre au sérieux,
c'est l'escalade.  

Rappelons deux caractéristiques de l'escalade.  Comme dans tous les jeux séquentiels,
seules comptent les comparaisons\,\footnote{De toutes façons, l'usage de probabilités, si
  toutefois on savait sur quoi les faire porter, n'ajouterait pas de prescription dans
  les choix successifs.}.  Donc les acteurs manipulent des entités qui sont complètement
abstraites: ils ne savent pas combien vaut ce qu'il manipulent, mais seulement comment les
entités se comparent.  D'autre part, les acteurs ayant, à chaque étape, un choix à faire
face à deux (ou plusieurs) options également rationnelles, l'évolution du processus est
fortement imprévisible. C'est à ce stade que peuvent intervenir des influences exogènes
dans le processus de décisions.  Puisqu'Alice n'a pas de raisons objectives de choisir
entre \og continuer\fg{} ou \og abandonner\fg{}, elle peut prendre en considération des
aspects plus émotionnels ou invoquer d'autres critères raisonnables.

\section{L'escalade et la psychologie cognitive}
\label{sec:esc-psy}
\hfill\parbox{12cm}{
 \begin{quote}
Il y a beaucoup de différence entre l'esprit de géométrie et l'esprit de finesse. 
\end{quote}
\rightline{\emph{Blaise Pascal,} \textsf{Pensées,} XXXI} }

\bigskip

Il est intéressant de se demander si les agents que nous étudions sont vraiment rationnels
ou plus précisément s'ils sont les plus rationnels possibles.  Prenons pour cela le point
de vue de la psychologie cognitive tel qu'il est décrit dans le livre de Keith
E. Stanovich \emph{What Intelligence Tests Miss: the psychology of rational
  thought}\,\footnote{\emph{Qu'est-ce qui échappe aux test d'intelligence? La psychologie
    de la pensée rationnelle}, \textsf{Yale University Press 2010}}.  Un agent rationnel
a à sa disposition un \emph{kit de raisonnement} (appelé en anglais un \emph{mindware)}
qui est constitué des règles, connaissances et procédures qu'il peut retrouver dans sa
mémoire, qu'il a acquises par un entraînement mental et/ou par l'éducation et qui lui permettent
de prendre des décisions et de résoudre des problèmes.  Bien sûr, nous supposons que le
kit de raisonnement des agents que nous considérons contient les outils du raisonnement
coinductif ou de tout type de raisonnement équivalent qui permet de conduire des
déductions correctes sur les objets mathématique infinis.

On distingue deux types de rationalité de la plus élementaire à la plus élaborée. D'une
part, la \emph{rationalité instrumentale} permet à l'agent de se comporter dans le monde
de telle façon qu'il obtienne ce qu'il désire le plus, compte tenu des ressources
(physiques et mentales) qui sont à sa disposition.   Les économistes et les
spécialistes de sciences cognitives ont raffiné la notion d'optimisation du but à
atteindre en celle d'\emph{utilité attendue}.  La \emph{rationalité épistémique}, quant à
elle, se place au dessus de la rationalité instrumentale et en interaction avec elle et
elle permet à l'agent de confronter l'ensemble de ses croyances à la structure effective
du monde. Nous dirions qu'elle fait prende à l'agent du recul, par rapport à son
raisonnement immédiat.  Plus prosaïquement, on peut dire que la rationalité épistémique
porte sur ce qui est vrai tandis que la rationalité instrumentale porte sur les actions à
faire pour maximiser ses objectifs.  La première forme de rationalité correspond à
l'esprit algorithmique et la deuxième à l'esprit réflexif.  Un esprit réflexif est capable
d'analyser la façon dont il raisonne.

On peut considérer qu'un agent escaladeur possède un esprit algorithmique correct, mais
manque d'esprit réflexif, ce qui lui  permettrait de changer ses croyances pour sortir de
la spirale de l'escalade.  En particulier, il devrait réviser sa croyance en des
ressources infinies.  Certes, au départ cela lui donne un certain dynamisme et ne l'inhibe
pas dans sa course en avant, mais comme dit l'adage \og Errare humanum est, perseverare
diabolicum\fg{}.  Il y a donc un moment où il faut comprendre que la croyance en cette
infinité de la ressource conduit à une impasse et où il faut reviser son jugement. En
faisant ainsi, suffisamment tôt l'agent montre qu'il est vraiment rationnel.

\section{L'ubiquité de l'escalade }
\label{sec:lubiq-de-lesc}
  \hfill\parbox{12cm}{
\begin{quote}
Elle s'acheta des plumes d'autruche, de la porcelaine chinoise et des bahuts ; elle
empruntait à Félicité, à Mme Lefrançois, à l'hôtelière de la Croix-Rouge, à tout le monde,
n'importe où. Avec l'argent qu'elle reçut enfin de Barneville, elle paya deux billets ;
les quinze cents autres francs s'écoulèrent. Elle s'engagea de nouveau, et toujours
ainsi ! 

\rightline{\emph{Gustave Flaubert}} \rightline{\textsf{Madame Bovary}}
\end{quote}}

\bigskip

L'escalade est un phénomène très courant dès que les participants sont rationnels et
considèrent que les ressources sont infinies.  Dans certains cas, c'est un objectif
recherché, car c'est la condition de survie, comme dans le cas de la biologie évolutive.
Dans sa théorie de la \emph{reine rouge}, Leigh Van Valen décrit la compétition de deux
espèces et la condition de survie de chaque espèce qui en résulte: à savoir une adaptation
continuelle pour lutter contre le défi de l'autre espèce.  L'escalade n'est donc pas un
défaut, mais une qualité positive de la pérennisation
de l'espèce quand elle a un compétiteur.  En revanche, dans le domaine économique, nous
savons très bien à quoi peut conduire l'escalade, à savoir des bulles spéculatives qui
engendrent de considérables dégâts quand elles éclatent. 

\begin{figure}[htb]
  \centering
  \begin{center}
    \includegraphics[width=.9\textwidth]{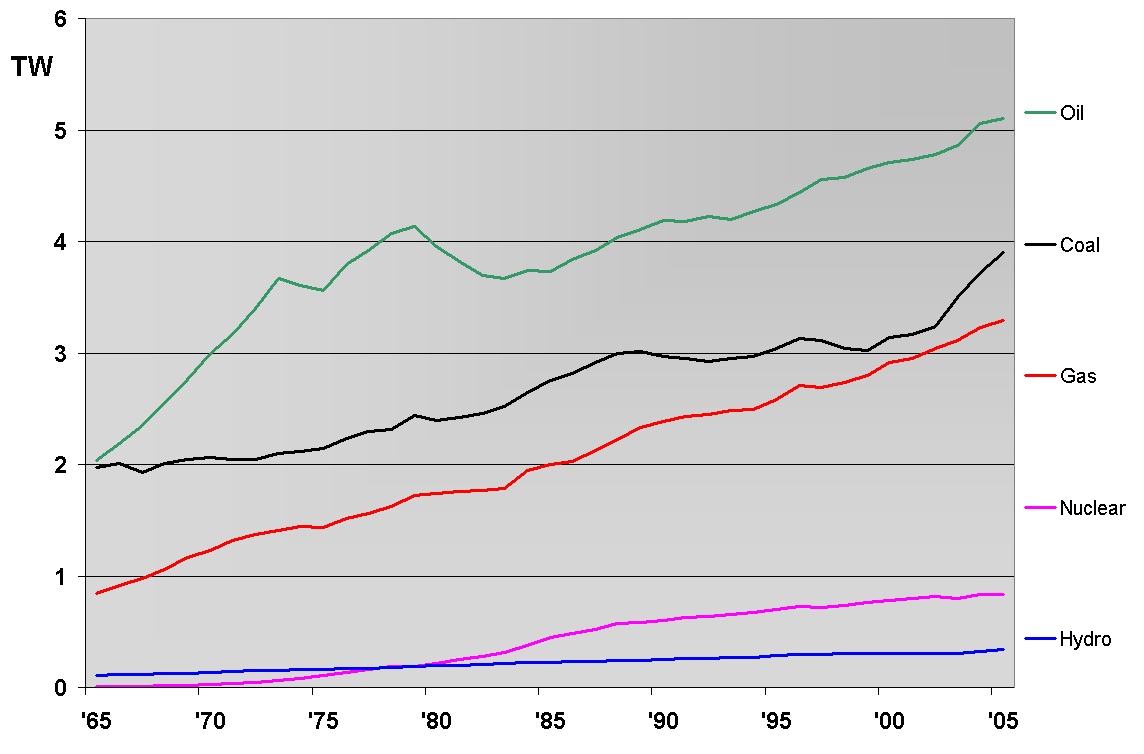}
  \end{center}
  
  \caption{Evolution de la consommation mondiale d'énergie (source Wikipédia)}
\label{fig:cons}
\end{figure}

Les théories du développement affirment que la terre, sous la forme d'énergies fossiles et
le soleil, par son rayonnement, nous fournit des ressources en quantité limitée et
donc nous incitent à nous méfier de la fuite en avant. La hausse de la consommation d'énergie
(Fig.~\ref{fig:cons}) est une inquiétante escalade.  

Dans un autre champ, l'escalade est une composante consubstantielle de la guerre.  Du
putsch de la Brasserie à son suicide en passant par l'écriture de \emph{Mein Kampf} et la
bataille de Stalingrad, la trajectoire d'Adolf Hitler est une escalade. Localement et
isolément, Hitler a fait preuve d'une rationalité stratégique.

\section{Il n'est pas possible d'extrapoler}
\label{sec:il-nest-pas}
\hfill\parbox{12cm}{
  \begin{quote}
    Un matin nous partons, le cerveau plein de flamme,

    Le c{\oe}ur gros de rancune et de désirs amers,

    Et nous allons, suivant le rythme de la lame,

    Berçant notre infini sur le fini des mers.

\centerline{\vdots}

Nous voulons, tant ce feu nous brûle le cerveau,

Plonger au fond du gouffre, Enfer ou Ciel, qu'importe ?

Au fond de l'Inconnu pour trouver du nouveau !

\rightline{\emph{Charles Baudelaire}}

\rightline{\textsf{Le voyage}}

  \end{quote}
}
\bigskip

\begin{figure}[hbp!] \label{fig:julia}
  \centering
  \includegraphics[width=.5\textwidth]{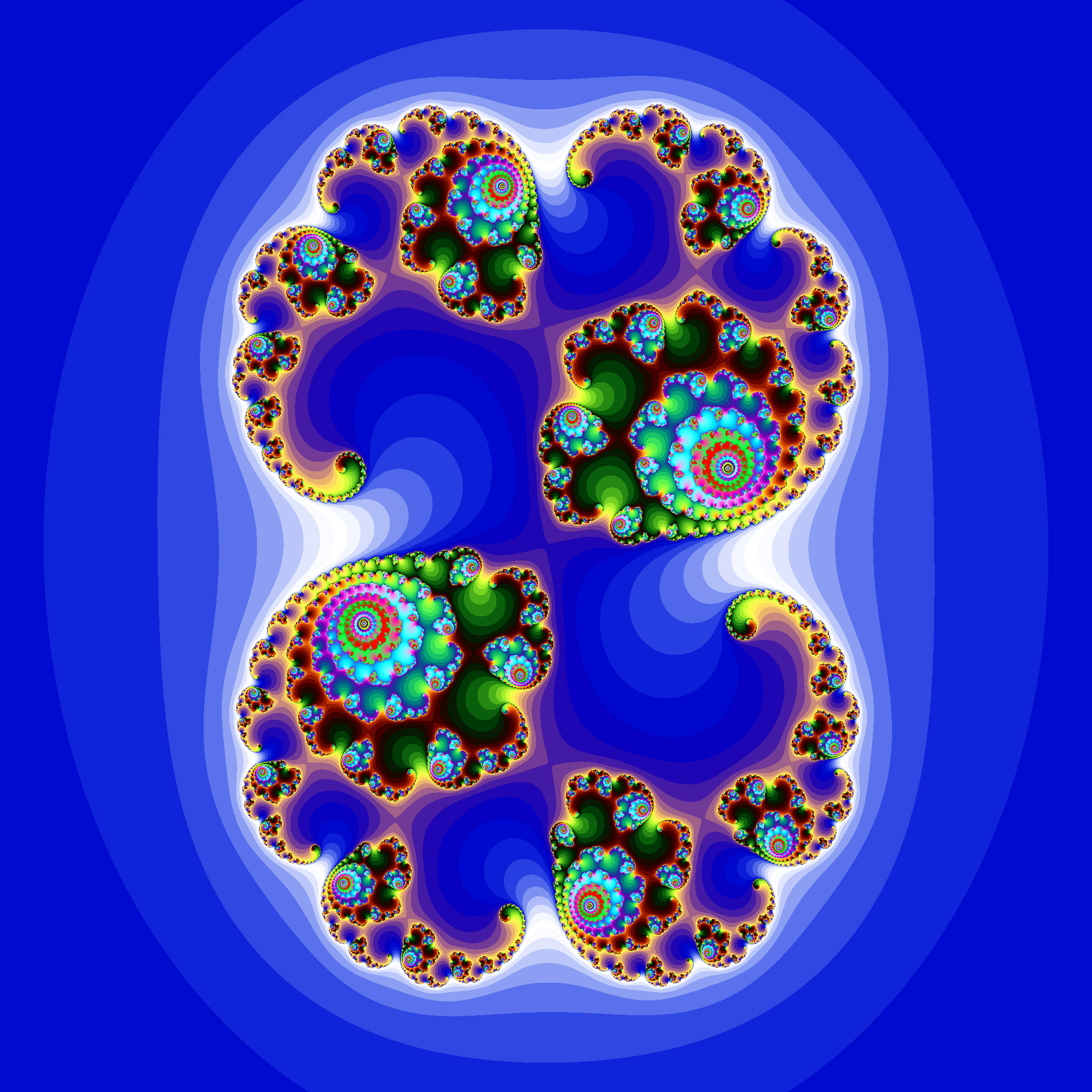}

\caption{L'ensemble de Julia, dont le contour est un courbe sans tangentes}
\end{figure}

Dans la section~\ref{sec:retour-sur-0,1}, nous avons vu que l'erreur du raisonnement de
Shubik et de ceux qui l'ont suivi consistait à \og élaguer\fg{} le jeu infini, puis à
extraire un jeu fini, puis à faire un raisonnement sur le jeu fini, puis à extrapoler ce
résultat au jeu infini pour en déduire une propriété.  Nous avons vu que sur le jeu $0,1$,
cette méthode ne peut absolument pas fonctionner: suivant que l'on élague à une étape paire
ou à une étape impaire, on obtient des résultats différents et l'on n'a aucune idée de
comment extrapoler.  Dans le cas de l'enchère à l'américaine, le problème est plus
vicieux, car les résultats semblent cohérents sur les différentes tailles de jeux finis,
encore que cela puisse dépendre de la façon dont on fait la coupe. Ce qui cloche dans
l'extrapolation, c'est un fait bien identifié depuis que le mathématicien Wei\-er\-strass
l'a mis en évidence en 1871: ce que l'on sait sur le fini ne préjuge en rien de ce que
l'on peut dire sur l'infini.  Plus précisément, il a montré qu'une propriété bien connue
et bien classique des sommes finies de fonctions (le fait d'être dérivables partout, les
courbes qui les représentent sont bien lisses), disparait pour les sommes infinies de
fonctions (qui peuvent être dérivables nulle part et les courbes qui les représentent être
particulièrement rugueuses).
Ce
résultat a surpris quand il a été publié, car de grands mathématiciens avant lui avaient
admis sans démonstration la persistance à l'infini de la dérivabilité.  De plus, cela
mettait en évidence l'existence de courbes monstrueuses, les courbes sans tangente nulle
part, qui, elles mêmes, préfiguraient les fractales de Mandelbrot (voir
Fig.~\ref{fig:julia}).
En fait, cette erreur dans l'extrapolation remonte à Zénon d'Élée, quand il affirme
qu'Achille ne pourra pas rattraper la tortue, ce qui conduit Zénon à nier le mouvement. En
effet, il extrapole le résultat vrai que, sur une course où Achille part derrière la
tortue et arrive à l'emplacement d'où la tortue est partie, Achille ne rattrape pas la
tortue. Il y a une infinité de telles courses.  Mais là où Zénon se trompe, c'est quand il
extrapole son résultat, vrai sur une course, à la limite de la suite infinie de ces
courses. Son raisonnement étant faux, il ne devrait donc pas pouvoir en conclure qu'Achille
ne rattrapera jamais la tortue.  

Considérons un autre exemple, à savoir un résultat mathématique récent et facile à
comprendre, car fondé sur des concepts connus de Pythagore et Euclide.  Une
\emph{progression arithmétique} est une suite obtenue en partant d'un certain nombre qui
sert d'origine et en ajoutant toujours le même nombre, appelé la raison\,\footnote{Ce
  concept de \og raison\fg{} n'a absolument rien à voir avec celui qui sous-tend la rationalité.}.  La suite
$5, 8, 11, 14, 17, ... $ est une suite arithmétique de raison $3$.  Il n'existe pas de
progression arithmétique infinie qui soit faite uniquement de nombres
premiers\,\footnote{Un nombre premier est un nombre qui est divisible seulement par le deux
  nombres $1$ et lui-même.  Si l'origine de la progression arithmétique est $n$ et sa
  raison est $r$, le $n+1^\textrm{ème}$ élément $n+ n\times r$ est clairement non premier,
  car divisible par $n$. }.  Mais Ben Green et Terence Tao ont démontré en 2004 qu'il
existe des progressions arithmétiques finies arbitrairement longues faites seulement de
nombres premiers et ce résultat très difficile a valu à Tao la médaille Fields.  On est
clairement en face d'un résultat qui n'est pas extrapolable.  L'intérêt tout particulier
de cet exemple est que le cas fini est incommensurablement plus difficile que le cas
infini.  Nous étions plus habitués au cas inverse, à savoir que le cas fini est beaucoup
plus facile que le cas infini qui nécessité la subtilité de la coinduction.

\section{Conclusion}
\label{sec:conclusion}

Par une analyse précise de l'infini, nous avons montré que les agents impliqués dans une
escalade sont (instrumentalement) rationnels, de leur point de vue.  Par conséquent, nous pensons que les
approches qui affirment un peu vite leur irrationalité\,\footnote{que nous ne nions pas,
  d'autant moins que, devant l'embarras du choix, ils peuvent invoquer des
  arguments irrationnels pour lever leur perplexité.} auraient dû prendre en compte l'analyse
coinductive des jeux infinis. L'utilisation de la coinduction et plus généralement des
raisonnements sur l'infini doivent donc faire partie des nouveaux fondements
des sciences économiques et de leurs outils ou comme disent les psychologues faire partie
de leur kit de raisonnement, leur mindware.  Dans ce cadre, équilibre ne rimera pas
forcément avec stabilité. 

D'autre part, la rationalité interne de l'agent, doit être nettement distinguée de la
rationalité externe de l'observateur.  L'agent, qui stipule une disponibilité infinie de
la ressource est un introverti qui ne voit que son seul intérêt à court terme, manque de
recul et raisonne sans \og réflexion\fg{}.  Il ne peut pas ou ne veut pas imaginer une
analyse globale, tandis l'observateur voit immédiatement l'aberration du comportement du
système.  Chacun a donc sa rationalité et les points de vue ne sont pas conciliables.
Cette réconciliation, si réconciliation il y a, n'est-elle pas la caractéristique de la
rationalité épistémique?


\pagebreak[4]

\end{document}